# Dynamic Control of Magnetically Trapped Indirect Excitons by Using External Magnetic Bias

Ahmed M. Abdelrahman and Byoung S. Ham[*]
*Photon Information Processing Center, School of Electrical Engineering,
Inha University, Incheon 402-751, South Korea*
* bham@inha.ac.kr

## Abstract

We demonstrate an on demand spatial control of excitonic magnetic lattices for the potential applications of excitonic-based quantum optical devices. A two dimensional magnetic lattice of indirect excitons can form a transition to one dimensional lattice configuration under the influence of external magnetic bias fields. The transition is identified by measuring the spatial distribution of two dimensional photoluminance for several values of the external magnetic bias fields. The number of the trapped excitons is found to increase between sites along a perpendicular direction exhibiting two to one dimensional lattice transition. This work may apply for various controllable quantum simulations, such as superfluid-Mott-insulators, in quantum optical devices.

PACS number (s): 71.35.Ji, 52.55.Jd, 52.55.Lf

## Introduction

Trapping of excitonic quasi-particles in solid state medium using electric or magnetic fields has recently become an active field of interest in the area of semiconductor physics. The two trapping mechanisms are rapidly gaining credits for their promises to underlie basic concepts of condensed matter systems. For example, it is now possible to create one and two dimensional lattice-like configurations of electrically trapped indirect excitons (bound state of electron-holes) in a system of coupled quantum wells (CQWs)[1-3]. Electrostatic traps were observed in cold excitonic gases, whose advantages are extending the lifetime of the trapped excitonic particles as well as allowing to observe the direct emission of their spontaneous coherence of the trapped indirect excitons[4]. In such approach the gate voltage controls the exciton's energy, where the applied electric field perpendicular to the growth direction of the CQWs shifts the energy of the created indirect excitons[5]. Moreover, the spatial distribution of the electric fields can configure a profile of confining potential for the excitonoic particles[5-7]. Accordingly, the electrical variable trapping field and its configurability made it possible to create *in situ* gate control for manipulating the trapped excitonic particles on a time scale that is shorter than their lifetimes[8].

On the other hand, magnetic field confinement has emerged recently as an alternative stable trapping mechanism that can be used to confine excitonic particles in solid mediums[9-13]. After its successful implementation in an atomic medium, magnetic traps have shown extremely stable trapping and a long coherence lifetime of the trapped particles[14]. These magnetic traps can also be configured to shape one and two dimensional magnetic lattices of the trapped particles[13,14,16]. The impeded magnetic trapping fields in semiconductors can be realized by integrating a fabricated permanent magnetic material, such as a form of magnetic thin film, with a system of CQWs[12-13]. The magnetic thin film can then be patterned according to a desirable spatial distribution of the magnetic trapping potentials. Similar to electrical traps, the shape and depth of the magnetic traps can be controlled by using external magnetic bias fields.

In this approach, excitonic particles are created in CQWs and trapped by magnetic fields representing two dimensional magnetic lattices of indirect excitons. This approach can be adopted to simulate condensed matter systems where strongly correlated systems such as the transition of superfluid-Mott-insulator (SF-MI) can be achieved[17]. In this article, we demonstrate a dimensional active control of the magnetic lattice configuration of indirect excitons by applying external magnetic bias fields.

## Two-Dimensional Magnetic Lattices

To create a magnetic lattice of indirect excitons, periodically distributed magnetic field local minima $B_{min}$ and maxima $B_{max}$ are projected onto the indirect-excitons formation plane of the

CQWs[13]. The inhomogeneous magnetic fields that used to trap the excitonic particles are originated from a specific pattern permanently magnetized thin film[16]. Figure 1(a) shows an example of two-dimensional arrays of the permanent magnetic square holes. The projected magnetic field local minima and maxima introduce periodic confining space (single traps) onto the plane of the CQWs, where the confinement mechanism acts a role of magnetic trapping of ultracold atoms[14-16].

An analytical model to describe the confining magnetic fields at a working distance $z_{min}$ from the surface of the thin film can be written as follow[16]:

$$B(x,y,z) = \Big[B_{x-bias}^2 + B_{y-bias}^2 + B_{z-bias}^2 + 2B_o^2 e^{-2\beta(z-\tau)}[1+\cos(\beta x)\cos(\beta y)] +$$
$$+2B_o^2 e^{-\beta(z-\tau)}[B_{x-bias}\sin(\beta x) + B_{y-bias}\sin(\beta y) + (\cos(\beta x)+\cos(\beta y))B_{z-bias}]\Big]^{1/2}$$
(1)

External magnetic bias fields are denoted by $B_{x,y,z-bias}$, and the thickness of the permanently magnetized thin film is characterized by the parameter τ, which is set to be equal to 2μm in both the simulation and the experiment. The $B_o$ is the reference magnetic field defined by $B_o = \tilde{B}(1-e^{-\beta\tau})$ with $\beta = \pi/\alpha$, where $\alpha$ represents both the length $\alpha_h$ of each fabricated pattern and their separating distances $\alpha_s$, as shown in Fig. 1(a). The $\tilde{B}$ is magnetic induction, $\tilde{B} = \frac{\mu_o M_z}{\pi}$, where $M_z$ is the magnetization of the thin film. The distribution of the magnetic lattice sites (single traps) is periodic according to Eq. (1), and the position of the local field minima $B_{min}$ are $x_{min} = n_x\alpha$, $n_x = 0, \pm 1, \pm 2, ...$, $y_{min} = n_y\alpha$, $n_y = 0, \pm 1, \pm 2, ...$, and $z_{min} \approx \frac{\alpha}{\pi}\ln[B_o]$.

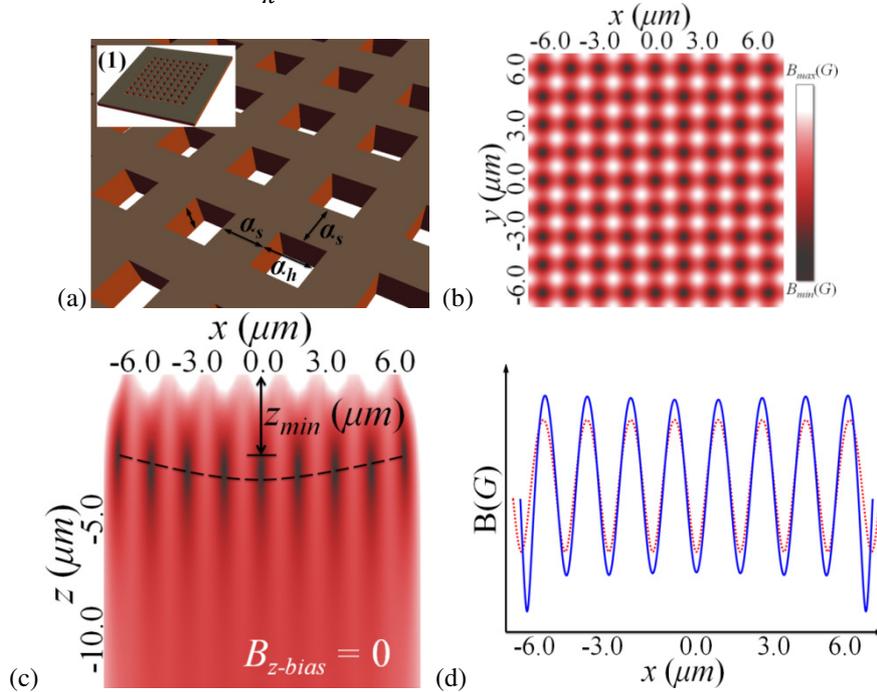

**Fig. 1.** (a) Simulation model of a permanent magnetic material thin film with thickness of τ=2μm used to generate the two-dimensional magnetic lattice (Inset (1) shows a block of 9×9 square holes). The lattice is created with periodicity of $\alpha_s = \alpha_h = \alpha = 2\mu m$. (b) Density plot of simulated lattice shows the distributed single traps across the x/y-plane at the local minima using analytical expression (Eq.1), and (c) numerically calculated trapping fields across the x/z-plane where the dashed line indicates the asymmetrical distribution of the sites along the z-axis. (d) Numerical (solid line) and the analytical (dashed line) calculations of the inhomogeneous magnetic field at a trapping level (i.e. z= $z_{min}(\mu m)$).

Figure 1(a) shows the model used to simulate the magnetic lattice, and Fig. 1(b) shows a magnetic lattice created at the plane of CQWs simulated by the analytical expression in Eq. (1). In the experimental setup each fabricated single pattern takes the shape of a square hole, where a set of n×n square holes is regarded as one block as shown in the inset of Fig. 1(a). The 3×3 blocks are fabricated, where each block consists of 9×9 square holes. As explained in the experimental details, the coupled quantum wells are allocated at $z = z_{min}(\mu m)$. However, due to the asymmetrical effect (that is the inhomogeneous spatial distribution of the lattice sites) across the x,y/z-plane, each confining area at the CQWs plane (i.e. at each individual lattice site) experiences different minimum value of trapping potential: This effect is indicated by the dashed line in Fig. 1(c). In other words, at zero bias fields the center sites have smaller values of the magnetic local minima $B_{min}(G)$ in comparison with the edge sites whose values are higher when measured at the confining plane of the coupled quantum wells (i.e., $B_{min}^{center}(G) < B_{min}^{edge}(G)$). The asymmetrical distribution of the sites creates periodically distributed potential tilts across the x/y-plane which is initially existed after magnetization[16]. The potential tilt is essential to allow the tunneling of trapped particles between the lattice sites, where the tunneling is the key mechanism of the present active control of the excitonic lattice transition.

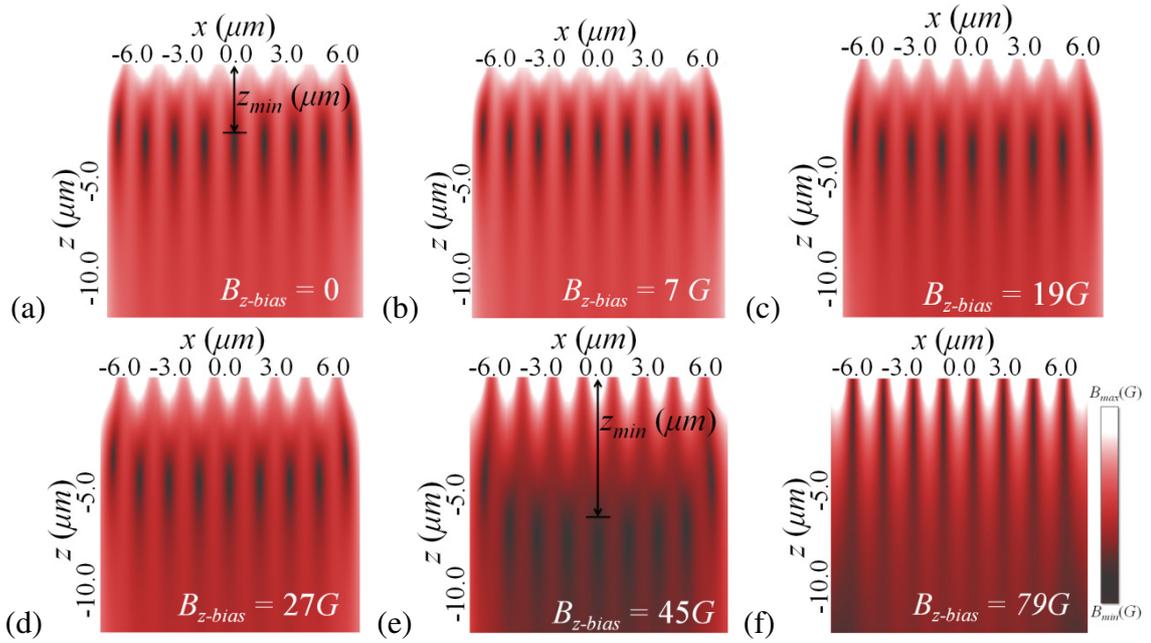

**Fig. 2.** Numerical simulations for the effect of external magnetic bias field along the z-axis. (a) The magnetic lattice is initially set at zero-bias: $B_{z-bias} = B_{x-bias} = B_{y-bias} = 0$); $\alpha_s = \alpha_h = 2\mu m$ $M_z = 2.8kG$; $\tau = 2\mu m$; n = 9 with local minima at $z_{min} \approx 4\mu m$. (b-e) Increasing z-bias field ($B_{x-bias} = B_{y-bias} = 0, B_{z-bias} \neq 0$) causes the magnetic local minima to depart away from the thin film surface deeper into and beyond the coupled quantum well plane. (f) A collapse of the local minima occurs at high values of $B_{z-bias}$ in which the magnetic lattice is reshaped as shown in Fig. 3(f).

The gradient (or the curvature) of the magnetic field at the trapping position is of a particular interest, where trapping field with steeper gradient and zero local minima (or very close to zero) often develops the so-called Majorana spin-flip in magnetic traps[18]. However, this destructive process has not yet been observed in such magnetic lattices of indirect excitons. It is important at this stage to emphasize that the zero local minima can easily be avoided in the case of magnetic trapping in semiconductors via fabrication; the non-zero local minima can precisely be allocated at the confining plane of the quantum wells.

The trapping frequencies $\omega_{x,y}$, along the x and y confining directions (that is across the x/y-plane of confinement) depend strongly on the gradient of the inhomogeneous magnetic fields at each site, where in our calculations we determine the curvatures of the trapping field along x/y-axis as follows:

$$\frac{\partial^2 B}{\partial x^2} = -\beta^2 B_o e^{\beta(z-\tau)} \left[ \frac{\cos(\beta x)\cos(\beta y)}{\sqrt{2+2\cos(\beta x)\cos(\beta y)}} + \frac{\cos^2(\beta x)\sin^2(\beta y)}{(2+2\cos(\beta x)\cos(\beta y))^{3/2}} \right], \quad (2)$$

$$\frac{\partial^2 B}{\partial y^2} = -\beta^2 B_o e^{\beta(z-\tau)} \left[ \frac{\cos(\beta x)\cos(\beta y)}{\sqrt{2+2\cos(\beta x)\cos(\beta y)}} + \frac{\cos^2(\beta y)\sin^2(\beta x)}{(2+2\cos(\beta x)\cos(\beta y))^{3/2}} \right]. \quad (3)$$

The trapping frequencies along x and y axes are defined as:

$$\omega_r = \frac{\beta}{2\pi} \sqrt{\mu_B g_F m_F \frac{\partial^2 B}{\partial r^2}} \quad \text{with } r \in \{x, y\}. \quad (4)$$

The trapping frequency along the z-axis is defined by $\omega_z = \omega_x + \omega_y$. In Eq. (4), $m_F$, $\mu_B$, and $g_F$ represent magnetic quantum number, the Bohr magneton, and the Lande g-factor, respectively. Each localized minimum is surrounded by magnetic barriers that define the space of the confinement and the depth of the traps, and the magnitude of the barrier is defined by $\Delta B(r) = |B_{max}(r)| - |B_{min}(r)|$, while the depth of each trap is defined by $\Gamma(r) = \frac{\mu_B g_F m_F}{k_B} \Delta B(r)$, where $K_B$ is the Boltzmann constant.

Figures 2-5 show the numerical simulation results of a two-dimensional magnetic lattice created with the following parameters: $\alpha_s = \alpha_h = 2\mu m$, $\tau = 2\mu m$, $M_z = 2.8KG$, and $n = 9$ holes. The working distance is found to be located at $z_{min} = (2\alpha \pm \Delta z)\mu m$, where the magnetic lattice is initially set at $B_{x-bias} = B_{y-bias} = B_{z-bias} = 0$ as shown in Fig. 2(a).

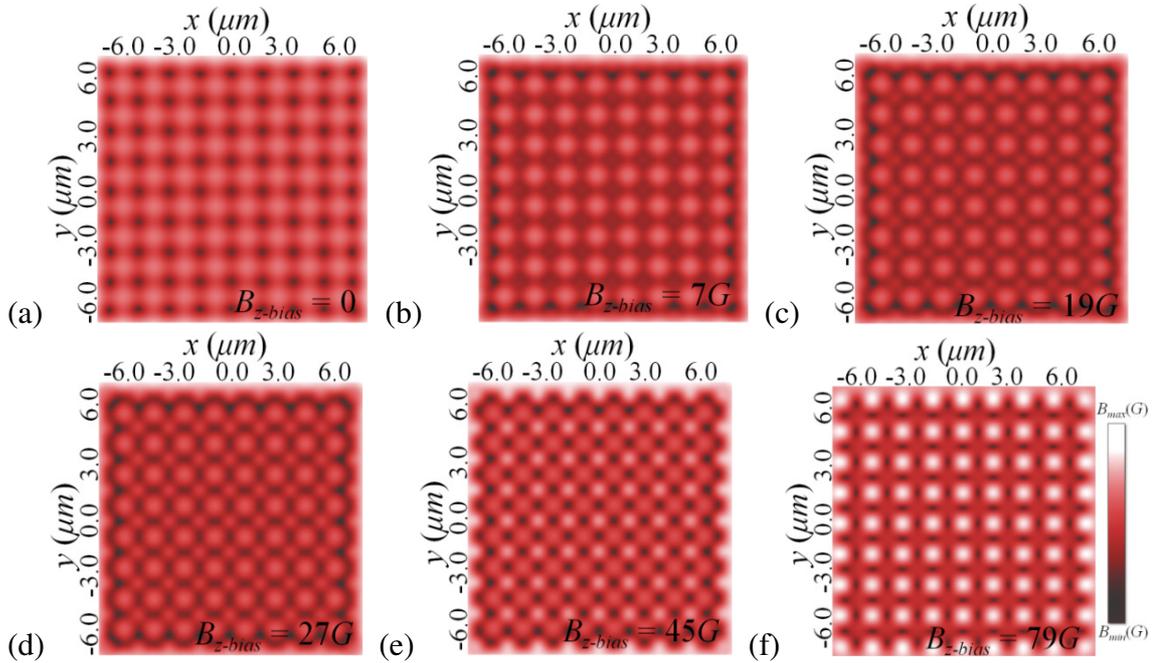

**Fig. 3.** Numerical calculations for magnetic field local minima distributed across the x/y-plane at $z_{min}$ with (a) no external magnetic bias field ($B_{z-bias} = B_{x-bias} = B_{y-bias} = 0$) for a lattice with the same initial parameter as in Fig. 2. (b-f) The shift of the lattice site locations and deformation in their confining spaces with respect to the external z-bias field: $B_{x-bias} = B_{y-bias} = 0, B_{z-bias} \neq 0$. All x/y-planes are simulated at the local minima along the z-axis: $z_{min} \approx 4\mu m$.

To induce the tunneling of the magnetically trapped particles or to displace the site locations across the x/y-plane and/or along the z-axis, the external magnetic bias fields are applied along the x/y-axis and/or the z-axis. The application of the bias fields, specifically $B_{z-bias}$ field, results in reducing the tunneling barrier between the sites and hence increasing the tunneling rate of the trapped excitons. The numerical simulation results shown in Figs. 2 and 3 are for the effect of the applied external bias field along the z-axis. The external magnetic z-bias field increases (or decreases) the distance $z_{min}$ between the surface of the permanently magnetized thin film and the initial position of the local field minima $B_{min}$, i.e., the actual location of the lattice site along the z-axis. As shown in Fig. 2(e), $B_{min/max}$ can be allocated at different distances from the plane of the coupled quantum wells in which case the magnetic field at the trapping point is increased (decreased) resulting in reduced (increased) number of trapped excitons. The z-bias field also causes to deform the confining space and to redistribute the lattice sites. As shown in Figs. 3(b)-(f), such effect allows the transition between different interesting configurations of the magnetic lattice.

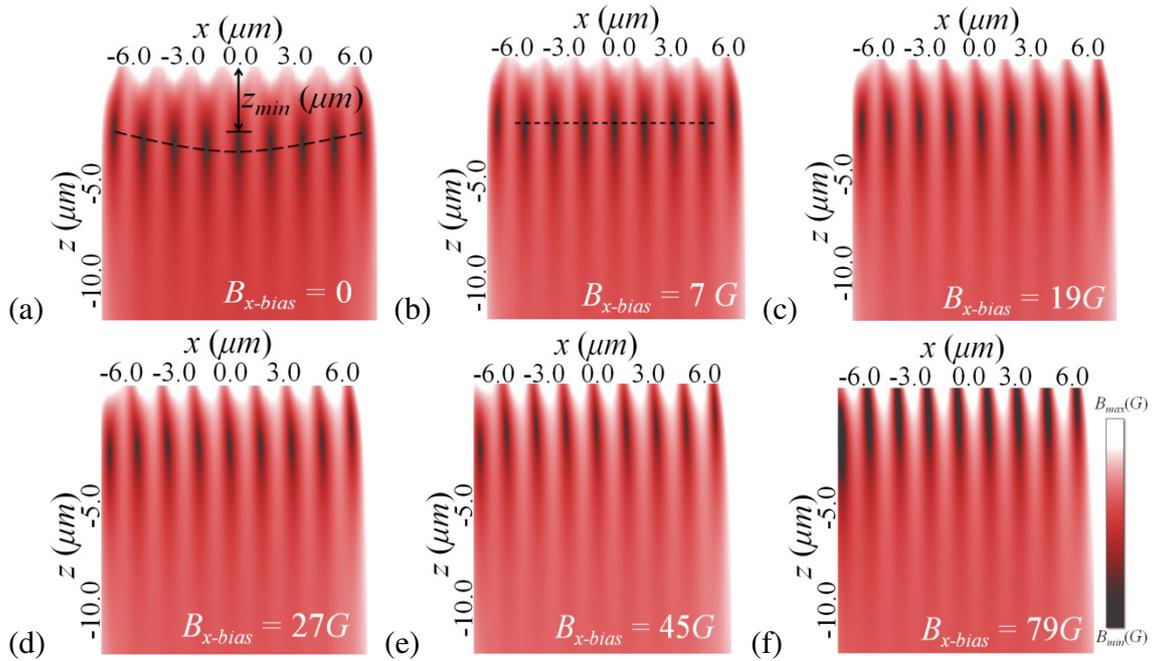

**Fig. 4.** (a) Using the same initial parameters of Fig. 2 a magnetic lattice subject to external magnetic bias field along the x-axis $B_{x-bias}$ is simulated: $B_{x-bias} = B_{y-bias} = B_{z-bias} = 0$. (b-f) The effects of the applied external bias field along the x-axis: $B_{y-bias} = B_{z-bias} = 0, B_{x-bias} \neq 0$. Oscillations between asymmetric (dashed line in (a)) and symmetric (dashed line in (b)) magnetic lattices can be controlled externally using the magnetic bias fields.

On the other hand, external magnetic bias fields along the x-axis and/or y-axis are used to reshape the magnetic lattices from two-dimension to one-dimension configuration. As shown in Figs. 5(b)-(f), external magnetic bias field along the x-axis can make the lattice sites to evolve from a two dimensional to a one dimensional distribution. A combination of z-axis and x/y-axis external magnetic bias fields may allow simulating condensed matter systems through reshaping the sites. For instance, it is possible to simulate the transition from/to Brillouin zones by redistributing the magnetically trapped excitonic particles across the lattice according to accurately adjusted external magnetic bias fields (this description also includes optically trapped atomic species[19]).

# Experimental Realization of On-demand 2D to 1D Lattice Transition in an Excitonic Magnetic Lattice

In our recent experimental results[13], we observed the formation of an excitonic two dimensional magnetic lattice at a plane of coupled quantum wells. For the present experiments, the CQWs are grown by molecular beam epitaxy, where there are two 8 nm GaAs quantum wells separated by a 4 nm $Al_{0.33}Ga_{0.67}As$ barrier. This CQWs is deposited by a 200 nm $Al_{0.33}Ga_{0.67}As$ conducting layer on both sides. Metal contacts are deposited at the top and at the bottom of the sample to monitor the electric field along the $z$-direction for generating indirect-excitons. A nonmagnetic material gadolinium gallium garnet $Gd_3Ga_5O_{12}$ (GGG) of thickness ≈3 μm is deposited, using an rf-sputtering technique, on top of the CQW system. The thickness of the GGG nonmagnetic spacer is used to determine the effective distance $z_{min}$ for allocating the magnetic field trapping local minima and maxima within the quantum well layers. The permanent magnetic material ($Bi_2Dy_1Fe_4Ga_1O_{12}$) with a thickness of ≈ 2 μm is deposited on top of the sample of the (GGG + CQWs) using rf-sputtering technique.

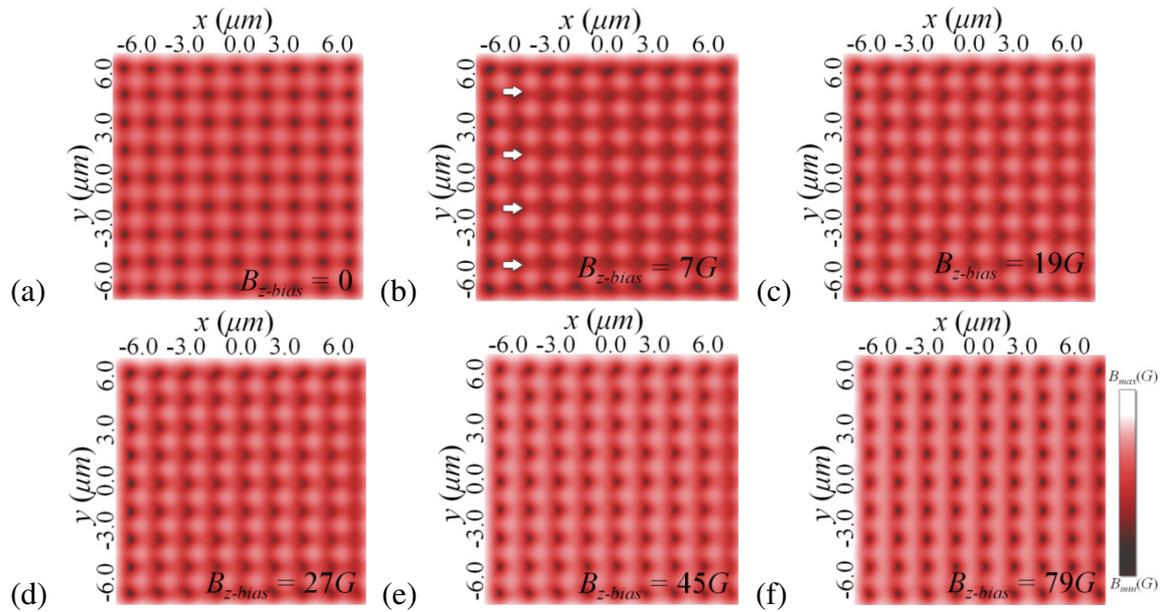

**Fig. 5.** External magnetic x-bias field-induced lattice sites across the x/y-plane. Sites in a two dimensional magnetic lattice can align themselves in specific patterns according to the external bias fields such as the transition to a one dimensional configuration (a-f). Initial parameters are the same as those in Fig. 2. (a) $B_{x-bias} = B_{y-bias} = B_{z-bias} = 0$. (b-f) $B_{y-bias} = B_{z-bias} = 0, B_{x-bias} \neq 0$ at $z_{min} \approx 4 \mu m$. Black arrows indicate the direction of the applied field.

The integrated magnetic-CQWs sample is placed in an optical cryostat with a temperature fixed at 13K, and the emitted PL images are collected using an objective microscope: For more details of the experimental methods see ref. 13. The indirect excitons are found to be self trapped according to the two dimensional distribution of the confining magnetic field, as shown in Fig. 6(a). As simulated in Fig. 5, external magnetic bias field along the x-axis is used to induce the bidirectional tunneling process between sites along the y-axis resulting in the observed 2D to 1D lattice transition as shown in Fig. 6. As shown in Fig. 6 the effect of gradually increasing the external magnetic bias field on the trapped indirect excitons is to reconfigure the shape of the magnetic lattice. Because the number of trapped particles per site changes according to the value of the externally applied magnetic field, the exciton redistribution in Fig. 6 strongly supports the theoretical calculations in Fig. 5.

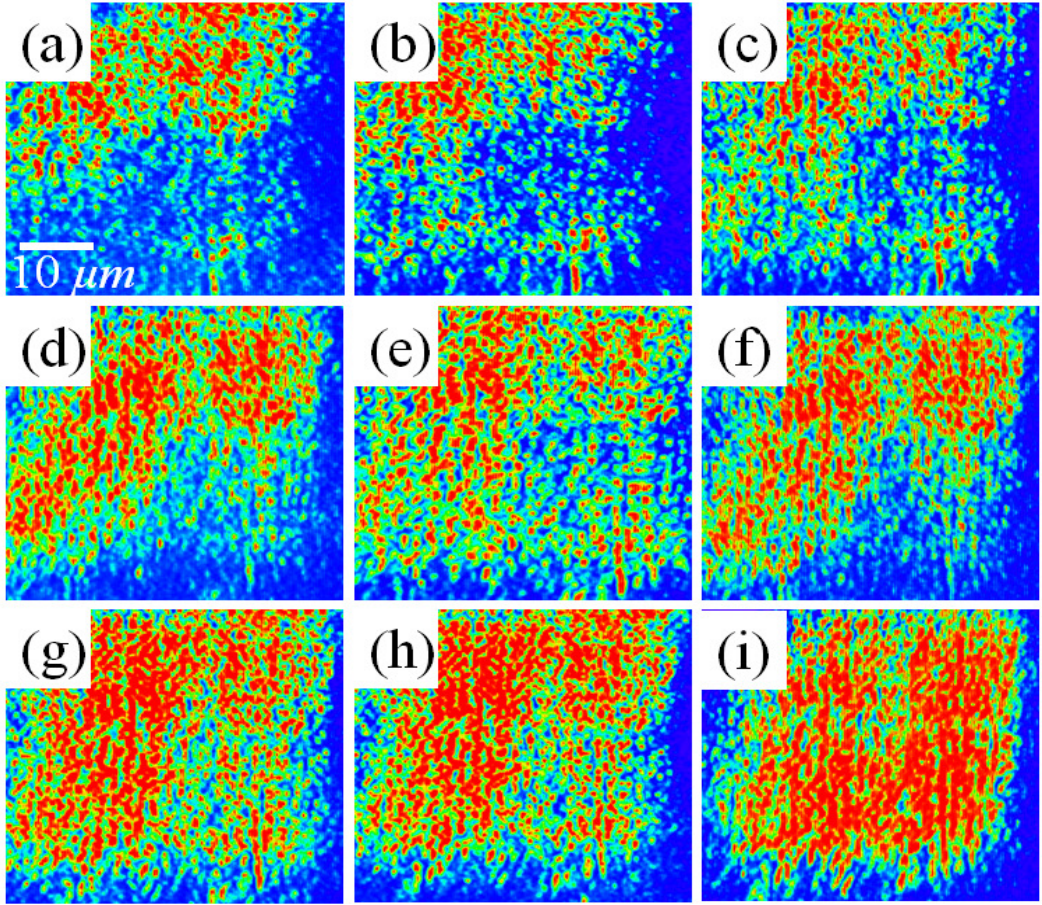

**Fig. 6.** Experimental results of the magnetically trapped indirect excitons showing 2D to 1D in lattice configuration. The magnetic lattice is created with $\alpha_s = \alpha_h = 2\mu m$  $M_z = 2.8 kG$, $\tau = 2\mu m$, and n = 9. The local filed minima projected at the coupled quantum wells level is $z_{min} \approx 2.5\ \mu m$. External magnetic bias field is applied along the x-axis and gradually increased from 0 to 75 G, $B_{y-bias} = B_{z-bias} = 0$, $B_{x-bias} = 0 \rightarrow 75G$. (a-h) Gate voltage is 0.2V. (i) Gate voltage increases to 0.3V to emphasize the excitonic concentration in 1D space.

Although the trapped excitonic gases are not sufficiently cooled, the transition effect in Fig. 6 is best interpreted in terms of a possibly reversed cycle of self-trapped excitonic clouds to the intra-sites superfluid state[20-23]. The reverse process can be achieved by decreasing the external magnetic bias fields (to be discussed elsewhere).

A scanning property of selected lattice sites along the y-axis in Fig. 7(a) shows increased number of trapped particles with the external magnetic x-bias field. Figure 7(b) shows the shift of trap positions along the x-axis with increased number of trapped excitons where, as a result, the external bias field causes a slight shift of the sites along the applied field direction, i.e., x-direction. In Fig. 8, trapped excitons are compared for two adjacent sites along the y-axis for different values of the external magnetic bias field applied along x-axis. The number of trapped particles between sites (in the valley along y-axis) increases as the value of the external bias field increases (along x-axis).

The results in Figs. 6-8 are in agreement with the simulation results of Fig. 5, a slight modification in the site's position and the shape along the x-axis of the lattice can be observed when applying external bias fields. The two to one dimensional lattice transition also occurs along the y-axis at the field bias constrains as simulated.

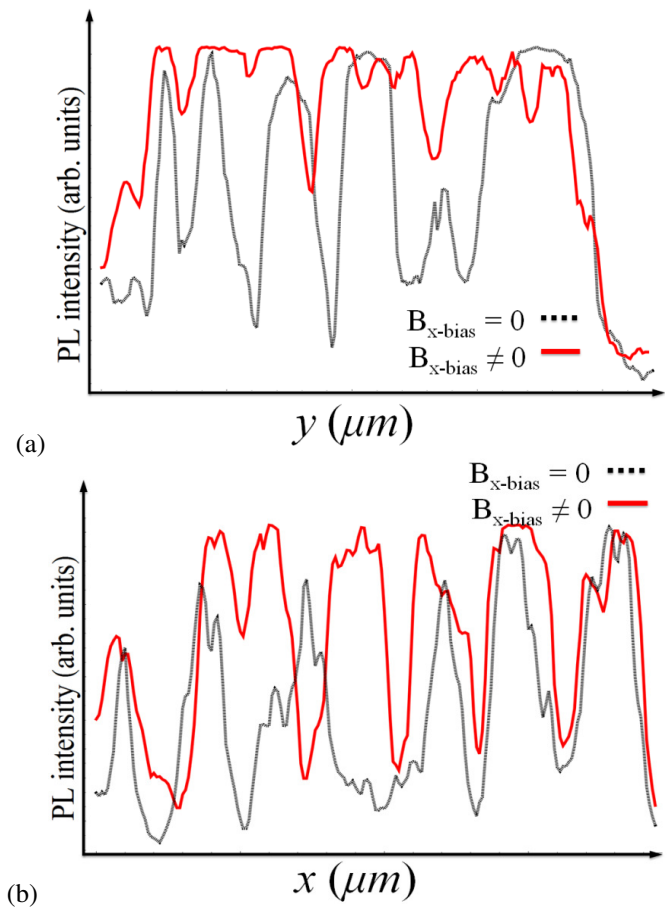

**Fig. 7.** (a) Measured PL intensity of the trapped indirect excitons across a selected number of lattice sites along the y-axis, before and after the external magnetic bias field. External x-bias field increases the number of trapped particles between lattice sites (along the y-axis) for 2D to 1D magnetic lattice transformation. (b) Measured PL intensity of the lattice sites along x-axis before and after the external x-bias field. The sites are also noticed to be displaced from their original positions with increased number of trapped particles. Experimental conditions are the same as in Fig. 6.

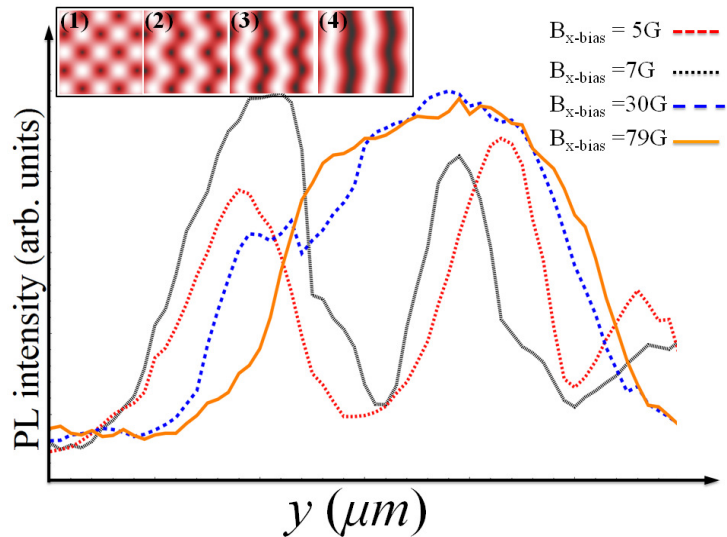

**Fig. 8.** PL measurements of two adjacent lattice sites along the y-axis for different external magnetic bias filed applied in x-axis. According to analytical calculations of the magnetic field in the insets (1)-(4) more indirect excitons are trapped between the sites because of the decreased magnetic field values between sites along y-axis. Experimental conditions are the same as in Fig. 6.

Emphasizing the fact that the current approach is suitable to simulate condensed matter systems, one can consider the magnetic confining field to act as the potential confining fields between ions, and the magnetically trapped excitonic particles play the role of electrons. In such a model the parameters are tunable without crystal-like disorders. The magnetic confinement approach can be applied to accommodate a strongly correlated system such as the transition of superfluid-Mott-insulator (MI) proposed by Jaksch et al.[17] and achieved in atomic medium by Greiner et al.[24-26]. Moreover, the preparation of the lattice in a MI state is a crucial step towards an efficient quantum register, where entanglement between the magnetically trapped excitons can be achieved by controlled collision using external magnetic bias fields[27].

## Conclusion

A dimensional active control of a magnetic lattice of excitonic particles was demonstrated. Magnetically trapped indirect excitons in a two dimensional lattice configuration can be controlled to transit into a one dimensional magnetic lattice of excitons with applied external magnetic bias fields. The dimentional transition is identified by measuring the spatial distribution of the 2D PL intensity for several values of the external bias fields where the number of the trapped excitons increases in the valley between sites, at a certain direction, exhibiting a 1D lattice configuration. In the current experiment, although the trapped excitonic gases were not sufficiently cooled to excitonic condensates (namely Bose-Einstein condensation of excitons), the observed transition can be interpreted in terms of self-trapped excitonic clouds to intra-sites superfluidity transition.

## Acknowledgment
This work was supported by the Creative Research Initiative Program (No. 2012-0000228) of the Korean Ministry of Education, Science and Technology via the National Research Foundation. BSH acknowledge that this work was also supported by the Korea Communications Commission, S. Korea, under the R&D program supervised by the Korea Communications Agency (KCA-2012-12-911-04-003).